\documentclass[aps,prd,floats,floatfix,twocolumn,nofootinbib,amssymb,amsmath]
{revtex4}

\usepackage{graphicx}
\usepackage{bm}

\newcommand{\beq}{\begin{equation}}
\newcommand{\eeq}{\end{equation}}
\newcommand{\beqa}{\begin{eqnarray}}
\newcommand{\eeqa}{\end{eqnarray}}

\begin{document}

\title{Optimal Calibration Accuracy for Gravitational Wave Detectors}

\author{Lee Lindblom}

\affiliation{
Theoretical Astrophysics 350-17, California Institute of
Technology, Pasadena, CA 91125}

\begin{abstract}
Calibration errors in the response function of a gravitational wave
detector degrade its ability to detect and then to measure the
properties of any detected signals.  This paper derives the needed
levels of calibration accuracy for each of these data-analysis tasks.
The levels derived here are optimal in the sense that lower accuracy
would result in missed detections and/or a loss of measurement
precision, while higher accuracy would be made irrelevant by the
intrinsic noise level of the detector.  Calibration errors affect the
data-analysis process in much the same way as errors in theoretical
waveform templates.  The optimal level of calibration accuracy is
expressed therefore as a joint limit on modeling and calibration
errors: increased accuracy in one reduces the accuracy requirement in
the other.
\end{abstract}


\date{\today}

\maketitle

\section{Introduction and Review}
\label{s:Introduction}

The response function is used to convert the electronic output of a
gravitational-wave detector into the measured gravitational-wave
signal.  This response function is determined experimentally by
performing a series of measurements when the detector is offline, and
then monitoring the output of the working, resonant detector (in a
time and frequency dependent way) as it reacts to inputs designed to
simulate its interaction with gravitational
waves~\cite{LIGOS4Calibration}.  The calibration procedure produces a
response function that is known therefore only to the level of
accuracy with which these various measurements are performed, and only
to the extent the state of the detector changes predictably between
calibration measurements.  This paper evaluates the effects of these
response-function errors on the subsequent gravitational-wave
data-analysis process, and from this determines the optimal levels for
calibration accuracy.

Inaccuracies in the response function degrade the ability to detect
gravitational-wave signals in the noisy data stream; and once
detected, they also reduce the ability to measure the physical
properties of the gravitational-wave source that produced the signal.
Errors in the gravitational-waveform models used in the data-analysis
process also degrade the detection and measurement procedures in a
very similar way.  An earlier discussion of the role of calibration
error on these data-analysis functions, cf. Ref.~\cite{Lindblom2008},
adopted the viewpoint that the level of calibration error was fixed.
The analysis there focused on determining the point at which further
reduction of waveform-modeling errors would be made irrelevant by the
presence of calibration error.  A more proactive viewpoint is adopted
here: that both the calibration error and the waveform-modeling error
levels can (in principle) be set to any desired level.  This paper
determines the optimal levels for the combined calibration and
waveform-modeling errors needed to perform detections and also to
perform measurements on any detected gravitational-wave signals.
These error levels are optimal in the sense that lower accuracy levels
would reduce the quantity and quality of the scientific information
extracted from the data; while higher accuracy would be made
irrelevant by the intrinsic noise level of the detector.

Let us begin by discussing briefly some of what is already known about
the effects of calibration error.  To that end, let us first establish
some notation.  Let $v(f)$ denote the direct electronic output of the
detector, and $R(f)$ the response function used to convert this raw
output to the inferred gravitational-wave signal $h(f)$:
\begin{eqnarray}
h(f)=R(f)v(f).
\end{eqnarray}
For simplicity, the discussion here is expressed in terms of the
frequency-domain representations of the various quantities.  For
example the frequency-domain waveform, $h(f)$, is related to its
time-domain analog, $h(t)$, by the Fourier transform:
\begin{eqnarray}
h(f)=\int_{-\infty}^{\infty} h(t)e^{-2\pi i f t} dt.
\end{eqnarray}
This transform follows the convention of the LIGO Scientific
Collaboration~\cite{T010095} (and the signal-processing community) by
using the phase factor $e^{-2\pi i f t}$, while most of the early
gravitational-wave literature and essentially all other computational
physics literature use $e^{2\pi i f t}$.  This choice does not affect
any of the subsequent equations in this paper.

Let us assume that the measured response function $R(f)$ differs from
the correct exact function $R_e(f)$ by $\delta R(f) =
R(f)-R_e(f)$.  This error in the response function will affect
measurements in two ways.  The response of the detector to a
gravitational-wave signal $h_e$ will produce an electronic output
$v_e$.  So the first effect of using the measured response function
$R$, is to interpret the signal as the waveform
$h=Rv_e=h_ee^{\delta\chi_R+i\delta\Phi_R}$, where the logarithmic
response function amplitude $\delta \chi_R$ and phase $\delta\Phi_R$
errors are defined by
\begin{eqnarray}
R= R_e+\delta R = R_e e^{\delta\chi_R+i\delta\Phi_R}.
\end{eqnarray}
This will produce a waveform error, 
\begin{eqnarray}
\delta
h_R=h_ee^{\delta\chi_R+i\delta\Phi_R}-h_e
\approx h_e(\delta\chi_R+i\delta\Phi_R), 
\end{eqnarray}
caused by the calibration error of the detector.  
The second effect of calibration error on measurements made with the
detector are errors in understanding the characteristics of the
detector noise.  In particular, the measured power spectral density of
the noise $S_n$ will differ from the exact $S_e$ due to the
calibration error $\delta R$.  The measured power spectral density of
the noise $S_n$ is related to $S_e$ by
\begin{eqnarray}
S_n(f) = S_e(f)\,e^{2\delta\chi_R}.
\label{e:MeasuredSh}
\end{eqnarray}

Both the detection and the measurement of a gravitational wave's
properties are adversely affected by response-function induced errors
in the waveform, $\delta h_R$, and the measured noise spectrum,
$S_n(f)$.  Similar adverse effects are caused by errors in the
waveform models used as part of the gravitational-wave data-analysis
procedure.  Let $\delta h_m(f) = h_m(f)-h_e(f)$ denote the difference
between a model gravitational waveform $h_m$, (e.g., one produced by a
numerical-relativity simulation) and the exact waveform $h_e$.  Both
types of waveform error, $\delta h_R$ and $\delta h_m$, cause
reductions in the signal-to-noise ratio, $\rho_m$, obtained when
a signal is projected onto the model waveform.  Keeping terms
through second-order in $\delta h_R$ and $\delta h_m$, it was shown 
previously~\cite{Lindblom2008} that the resulting
measured signal-to-noise ratio, $\rho_m$, is related to the optimal
signal-to-noise ratio, $\rho$, by the expression:
\begin{eqnarray}
\rho_m = \rho-\frac{1}{2\rho}\langle(\delta h_m-\delta h_R)_\perp|
(\delta h_m-\delta h_R)_\perp\rangle,
\label{e:modelpluscalibrationerror}
\end{eqnarray}
where the quantity $(\delta h_m - \delta h_R)_\perp$ is the projection of
$\delta h_m - \delta h_R$ orthogonal to the exact waveform,
\begin{eqnarray}
(\delta h_m - \delta h_R)_\perp &=&
\delta h_m -\delta h_R - h_e \frac{\langle \delta h_m - \delta h_R|h_e\rangle}
{\langle h_e|h_e\rangle}.\quad
\label{e:delta_h_perp_def}
\end{eqnarray}
The noise-weighted inner products, e.g., $\langle \delta h_m| \delta
h_R\rangle$, used in these expressions are defined with respect to the
measured power spectral density of the noise $S_n(f)$:
\begin{eqnarray}
&&\!\!\!\!\!\!\!\!\!\!\!\!\!\!\!\!\!\!\!\!
\langle \delta h_m| \delta h_R\rangle=\nonumber\\
&&\quad2\int_{0}^\infty \frac{
\delta h_m(f)\delta h^*_R(f)+\delta h_m^*(f)\delta h_R(f)}{S_n(f)}df.
\end{eqnarray}
The derivations of these expressions are given in some detail in
Sec.~III of Ref.~\cite{Lindblom2008}.

\section{Calibration Accuracy Standards}
\label{s:CalibrationAccuracyStandards}

The expression for the difference between the measured and optimal
signal-to-noise ratios $\delta\rho=\rho_m-\rho$ in
Eq.~(\ref{e:modelpluscalibrationerror}) is remarkably simple,
depending only on the difference between the waveform errors,
$\delta\rho =\delta\rho(\delta h_m-\delta h_R)$.  At the most basic
level, the waveform-accuracy standards developed in
Ref.~\cite{Lindblom2008} were obtained by limiting the size of the
waveform errors to those producing acceptably small changes in
$\delta\rho$.  Since $\delta\rho$ depends only on the difference in
waveform errors, $\delta h_m-\delta h_R$, from
Eq.~(\ref{e:modelpluscalibrationerror}), it follows that the
ideal-detector waveform accuracy standards can be extended to the
realistic-detector case ($\delta h_R\neq 0$) simply by replacing
$\delta h_m$ with $\delta h_m-\delta h_R$ in those ideal-detector
standards.  Thus the optimal accuracy requirement on the combined
(calibration plus modeling) waveform errors that ensures no loss of
scientific information during the measurement process is,
\begin{eqnarray}
\langle \delta h_m-\delta h_R | \delta h_m-\delta h_R\rangle < 1.
\label{e:measurmentoptimal}
\end{eqnarray}
This is the generalization of the ideal-detector condition derived as
Eq.~(5) of Ref.~\cite{Lindblom2008}.  Similarly the optimal accuracy
requirement on the combined waveform errors that ensures no
significant reduction in the rate of detections is,
\begin{eqnarray}
\langle (\delta h_m-\delta h_R)_\perp | (\delta h_m-\delta
h_R)_\perp\rangle < 2\rho^2\epsilon_\mathrm{max}.
\end{eqnarray}
The parameter $\epsilon_\mathrm{max}$ determines the fraction of
detections that will be missed as a result of calibration and modeling
errors, as discussed in some detail in Ref.~\cite{Lindblom2008}. As in
the ideal-detector case, it is more convenient to convert this optimal
accuracy requirement for detection into the slightly stronger
sufficient condition,
\begin{eqnarray}
\langle \delta h_m-\delta h_R | \delta h_m-\delta
h_R\rangle < 2\rho^2\epsilon_\mathrm{max},
\label{e:detctionbasic}
\end{eqnarray}
which does not require a knowledge of the projection $(\delta
h_m-\delta h_R)_\perp$.  This simpler expression is the generalization
of the ideal-detector condition derived as Eq.~(15) of
Ref.~\cite{Lindblom2008}.

Both of these accuracy standards, Eqs.~(\ref{e:measurmentoptimal}) and
(\ref{e:detctionbasic}), on the combined (calibration plus modeling)
waveform errors are conditions on the noise-weighted norm of $\delta
h_m-\delta h_R$.  The waveform-modeling errors, $\delta h_m$, have an
entirely different source and are therefore completely uncorrelated
with the calibration errors, $\delta h_R$.  It is useful therefore to
express these accuracy standards in a form that isolates each type of
error.  This can be done with a simple application of the Schwarz
inequality, cf. Eq.~(44) of Ref.~\cite{Lindblom2008}:
\begin{eqnarray}
&&\langle \delta h_m - \delta h_R | \delta h_m - \delta h_R\rangle
\leq \nonumber\\
&&\qquad\qquad\qquad
\left(\sqrt{\langle\delta h_m|\delta h_m\rangle}
+\sqrt{\langle\delta h_R|\delta h_R\rangle}\right)^2.
\qquad
\end{eqnarray}
This inequality is fairly tight, in the sense that equality is
actually achieved for the case $\delta h_m = - \delta h_R$.  Based on
this inequality, the following slightly stronger, sufficient, versions
of the accuracy requirements can be constructed for measurement,
\begin{eqnarray}
\sqrt{\langle \delta h_m| \delta h_m\rangle}
+\sqrt{\langle \delta h_R| \delta h_R\rangle}
 < 1,
\label{e:measurementlimit}
\end{eqnarray}
and for detection,
\begin{eqnarray}
\sqrt{\langle \delta h_m| \delta h_m\rangle}
+\sqrt{\langle \delta h_R| \delta h_R\rangle}
 < \sqrt{2\epsilon_\mathrm{max}}\,\rho.
\label{e:detectionlimit}
\end{eqnarray}
These conditions reduce to the model-waveform accuracy standards
derived in Ref.~\cite{Lindblom2008} for the ideal-detector case
($\delta h_R=0$).  These accuracy standards place more stringent
conditions, however, on the waveform-modeling error when there is a
non-negligible level of calibration error.

The allowed error levels, due to calibration and waveform
modeling, can be apportioned between the two error sources in any way
that is consistent with Eqs.~(\ref{e:measurementlimit}) and
(\ref{e:detectionlimit}).  Determining the most efficient way to do
this would require an analysis of the relative costs of improving the
accuracies of each error source.  It seems likely that adopting
standards with comparable requirements for each type of error will be
close to optimal.  Let us explore in some detail then what the
resulting calibration and modeling accuracy requirements would be in
that case.  From Eqs.~(\ref{e:measurementlimit}) and
(\ref{e:detectionlimit}) it follows that the appropriate limits 
become
\begin{eqnarray}
\langle \delta h_m| \delta h_m\rangle=
\langle \delta h_R| \delta h_R\rangle
 < \frac{1}{4},
\label{e:measurementlimit1}
\end{eqnarray}
for measurement, and
\begin{eqnarray}
\langle \delta h_m| \delta h_m\rangle=
\langle \delta h_R| \delta h_R\rangle
 < \frac{\epsilon_\mathrm{max}}{2}\rho^2,
\label{e:detectionlimit1}
\end{eqnarray}
for detection.  These waveform-modeling standards are stricter by a
factor of two than those derived in Ref.~\cite{Lindblom2008} for the
ideal-detector case.

It is useful to translate these accuracy requirements into a more familiar
language, by noting that (to lowest order) the waveform error can
be written in terms of logarithmic amplitude and phase errors:
$\delta h_m \approx h_e(\delta \chi_m+i\delta\Phi_m)$.
It follows that the norm of the waveform-modeling error can be
expressed in the form,
\begin{eqnarray}
\langle \delta h_m|\delta h_m\rangle =  
\rho^2\left(\overline{\delta\chi_m}^{\,2} +\overline{ \delta\Phi_m}^{\,2}
\right),
\end{eqnarray}
where the signal and noise weighted averages of the amplitude
and phase errors are defined by,
\begin{eqnarray}
\overline{\delta\chi_m}^{\,2}&=&\int_{0}^\infty 
\left(\delta\chi_m\right)^2
\frac{4|h_e|^2}{\rho^2S_n(f)}df,\\
\overline{\delta\Phi_m}^{\,2}&=&\int_{0}^\infty 
\left(\delta\Phi_m\right)^2
\frac{4|h_e|^2}{\rho^2S_n(f)}df.
\end{eqnarray}
Note that the weight term, $4|h_e|^2/\rho^2S_n$, which appears in
these definitions has integral one; so these are true (signal and
noise weighted) averages of $\delta \chi_m$ and $\delta \Phi_m$.  The
averages of the calibration amplitude and phase errors, are defined
analogously.  In terms of these averages then, the waveform-accuracy
standards of Eqs.~(\ref{e:measurementlimit1}) and
(\ref{e:detectionlimit1}) become
\begin{eqnarray}
\sqrt{\overline{\delta\chi_m}^{\,2} +\overline{ \delta\Phi_m}^{\,2}}=
\sqrt{\overline{\delta\chi_R}^{\,2} +\overline{ \delta\Phi_R}^{\,2}}
 < \frac{1}{2\rho},
\label{e:measurementlimit2}
\end{eqnarray}
for measurement and
\begin{eqnarray}
\sqrt{\overline{\delta\chi_m}^{\,2} +\overline{ \delta\Phi_m}^{\,2}}=
\sqrt{\overline{\delta\chi_R}^{\,2} +\overline{ \delta\Phi_R}^{\,2}}
 < \sqrt{\frac{\epsilon_\mathrm{max}}{2}},
\label{e:detectionlimit2}
\end{eqnarray}
for detection.  

For Advanced LIGO the maximum signal-to-noise ratio for a binary
black-hole signal may be as large as about 100, so the resulting
accuracy requirements sufficient for measurement, from
Eq.~(\ref{e:measurementlimit2}), for such events are
\begin{eqnarray}
\sqrt{\overline{\delta\chi_m}^{\,2} +\overline{ \delta\Phi_m}^{\,2}}=
\sqrt{\overline{\delta\chi_R}^{\,2} +\overline{ \delta\Phi_R}^{\,2}}
 \lesssim 0.005.
\label{e:measurementLIGO}
\end{eqnarray}
Thus the averages of the frequency-domain amplitude and phase errors
must be at about the 0.35\% and the 0.0035 radian levels respectively
for measurement.  If the Advanced LIGO search template banks are
constructed (as in Initial LIGO) with waveform templates spaced so
that no point in the template subspace has a mismatch larger than 0.03
from some element in the bank, then the maximum mismatch
$\epsilon_\mathrm{max}$ must be chosen to be 0.005 to ensure the
signal loss rate is about 10\%, cf. Ref.~\cite{Lindblom2008}.  In this
case the resulting accuracy requirements for both waveform and
calibration errors sufficient for detection from
Eq.~(\ref{e:detectionlimit2}) are
\begin{eqnarray}
\sqrt{\overline{\delta\chi_m}^{\,2} +\overline{ \delta\Phi_m}^{\,2}}=
\sqrt{\overline{\delta\chi_R}^{\,2} +\overline{ \delta\Phi_R}^{\,2}}
 \lesssim 0.05.
\label{e:detectionLIGO}
\end{eqnarray}
Thus the accuracy requirements for detection are an order of magnitude
less stringent than those needed for measurement of the strongest
likely sources in Advanced LIGO.  The required averages of the
frequency-domain amplitude and phase errors must be at about the 3.5\%
and the 0.035 radian levels respectively for detection.

It is easy to imagine how two different sets of model waveforms could
be designed to accomplish the two distinct data-analysis tasks.  One
set could be prepared for use in searches of gravitational-wave
signals using the lower accuracy standards needed for detection.  And
a second set could be prepared with the higher accuracy standards
needed for measurement, but only in the very small portion of waveform
parameter space where they are needed for measurements on previously
detected signals.  At present it seems unlikely that it will be
possible to perform detector calibrations in a way that provides the
lower level of calibration accuracy needed for detections at all
times, and only calibrates to the higher accuracy standards needed for
measurements {\it{ex post facto}} in those data segments where
detections have been made.  So it seems likely that it will be
necessary to calibrate the detectors at the level needed for
measurements, e.g., according to the standards of
Eq.~(\ref{e:measurementlimit2}), whenever data is collected.  In this
case, the accuracy standard for detections for waveform-modeling error
could be relaxed somewhat to the level
\begin{eqnarray}
\sqrt{\overline{\delta\chi_m}^{\,2} +\overline{ \delta\Phi_m}^{\,2}}
 < \sqrt{2\epsilon_\mathrm{max}}-\frac{1}{2\rho_\mathrm{max}}\approx 
0.095,
\label{e:detectionLIGO1}
\end{eqnarray}
which is almost identical to the ideal-detector requirement derived in
Ref.~\cite{Lindblom2008}.

\section{User Friendly Standards}
\label{s:UserFriendlyStandards}

The accuracy standards derived in
Sec.~\ref{s:CalibrationAccuracyStandards} for the combined
(calibration plus modeling) waveform errors are not easily applied
using the basic expressions given in Eqs.~(\ref{e:measurementlimit2})
and (\ref{e:detectionlimit2}).  Model waveforms are generally
constructed in the time domain (e.g., by performing numerical
simulations), so verifying the basic frequency-domain standards using
estimates of the time-domain errors is not
straightforward~\cite{Lindblom2009b}.  Neither can the basic
expressions for the standards on the calibration errors be enforced in
a straightforward way.  While good estimates of the frequency-domain
response-function errors are generally
available~\cite{LIGOS4Calibration}, the accuracy standards of
Eqs.~(\ref{e:measurementlimit2}) and (\ref{e:detectionlimit2}) require
computing their averages weighted by the gravitational waveform
$h_e$.  What waveform should be used when applying these standards?
This section transforms the basic accuracy standards of
Eqs.~(\ref{e:measurementlimit}) and (\ref{e:detectionlimit}) into
forms that are more easily used by those responsible for calibrating
the detector, and by those responsible for constructing and verifying
the accuracy of model waveforms as well.

\begin{figure}
\centerline{\includegraphics[width=3in]{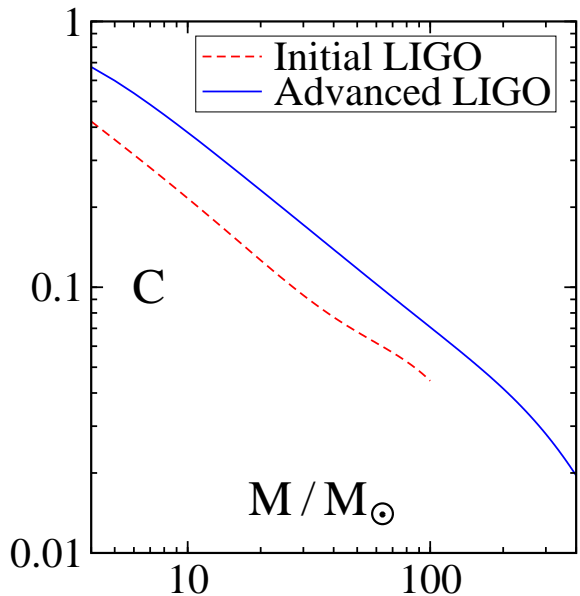}}
\caption{\label{f:CSNratio} Curves illustrate $C$, the ratio of the
standard signal-to-noise measure $\rho$ to a non-standard measure
defined in Eq.~(\ref{e:SNRatioRatio}), as a function of the total mass
for non-spinning equal-mass binary black-hole waveforms.  Dashed curve
is based on the Initial LIGO noise spectrum~\cite{InitialLIGONoise};
solid curve is based on an Advanced LIGO noise
curve~\cite{AdvancedLIGONoise}.}
\end{figure}
The norm of the waveform-modeling error $\langle \delta h_m|\delta
h_m\rangle$, which appears in the accuracy standards of
Eqs.~(\ref{e:measurementlimit}) and (\ref{e:detectionlimit}), is
constructed from the frequency-domain estimates of those errors.  It
is not straightforward to determine useful estimates of these
frequency-domain errors from the time-domain waveform errors that are
directly accessible to the waveform-modeling community.  It is
useful therefore to transform the expression for the limits on the
modeling error into ones based directly on time-domain estimates of
the errors.  This can be done, following the argument in Sec.~II.C of
Ref.~\cite{Lindblom2008}, using an application of Parseval's theorem:
\begin{eqnarray}
\langle\delta h_m|\delta h_m\rangle \leq
\frac{\rho^2}{C^{\,2}}\frac{||\delta h_m(t)||^2}{||h_e(t)||^2},
\label{e:time-domain-limit}
\end{eqnarray}
where $||\delta h_m(t)||$ is the $L^2$ norm of $\delta h_m(t)$, defined
as
\begin{eqnarray}
||\delta h_m(t)||^2=\int_{-\infty}^\infty |\delta h_m(t)|^2 dt,
\end{eqnarray}
and where $C$ is the ratio of the standard signal-to-noise measure
$\rho$ to a non-standard measure:
\begin{eqnarray} 
C^{\,2}=\rho^2\left(\frac{2||h_e(t)||^2}{\mathrm{min}\, S_n(f)}\right)^{-1}.
\label{e:SNRatioRatio}
\end{eqnarray}
Figure~\ref{f:CSNratio} illustrates $C$ for non-spinning equal-mass
binary black-hole waveforms (cf. Fig.~4 of Ref.~\cite{Lindblom2008}).
This quantity can be evaluated in a straightforward way when any class
of model waveforms is computed.  The right side of
Eq.~(\ref{e:time-domain-limit}) is always larger than the
noise-weighted norm $\langle \delta h_m|\delta h_m\rangle$ that
appears on the left.  Sufficient conditions for model-waveform
accuracy based on the time-domain $L^2$ norm $||\delta h_m(t)||$ can
be obtained therefore by replacing $\langle \delta h_m|\delta
h_m\rangle$ with the right side of Eq.~(\ref{e:time-domain-limit})
wherever it appears in the accuracy standards of
Eqs.~(\ref{e:measurementlimit}) and (\ref{e:detectionlimit}).

The norm of the waveform error caused by detector calibration errors,
$\langle \delta h_R|\delta h_R\rangle$, is also difficult to evaluate
because it depends on the gravitational waveform $h_e$ in addition to
the purely detector-based errors $\delta \chi_R$ and $\delta \Phi_R$.
The detector calibration-error terms in this norm can be isolated
from the gravitational-waveform terms in a straightforward way 
using the inequality
\begin{eqnarray}
\langle \delta h_R | \delta h_R\rangle &=&
\int_0^\infty \left(\delta\chi_R^2+\delta\Phi_R^2\right)
\frac{4|h_e|^2}{S_n}df,\nonumber\\
&\leq& \rho^2\,\mathrm{max}\left(\delta\chi_R^2+\delta\Phi_R^2\right).
\label{e:calibrationerror3}
\end{eqnarray}
The right side of Eq.~(\ref{e:calibrationerror3}) is very easy to
evaluate, and is always larger than the noise-weighted norm $\langle
\delta h_R|\delta h_R\rangle$ that appears on the left.  Sufficient
conditions for calibration accuracy based on
$\mathrm{max}\left(\delta\chi_R^2+\delta\Phi_R^2\right)$ can be
obtained therefore by replacing $\langle \delta h_R|\delta h_R\rangle$
with the right side of Eq.~(\ref{e:calibrationerror3}) wherever it
appears in the accuracy standards of Eqs.~(\ref{e:measurementlimit})
and (\ref{e:detectionlimit}).

\begin{figure}
\centerline{\includegraphics[width=3in]{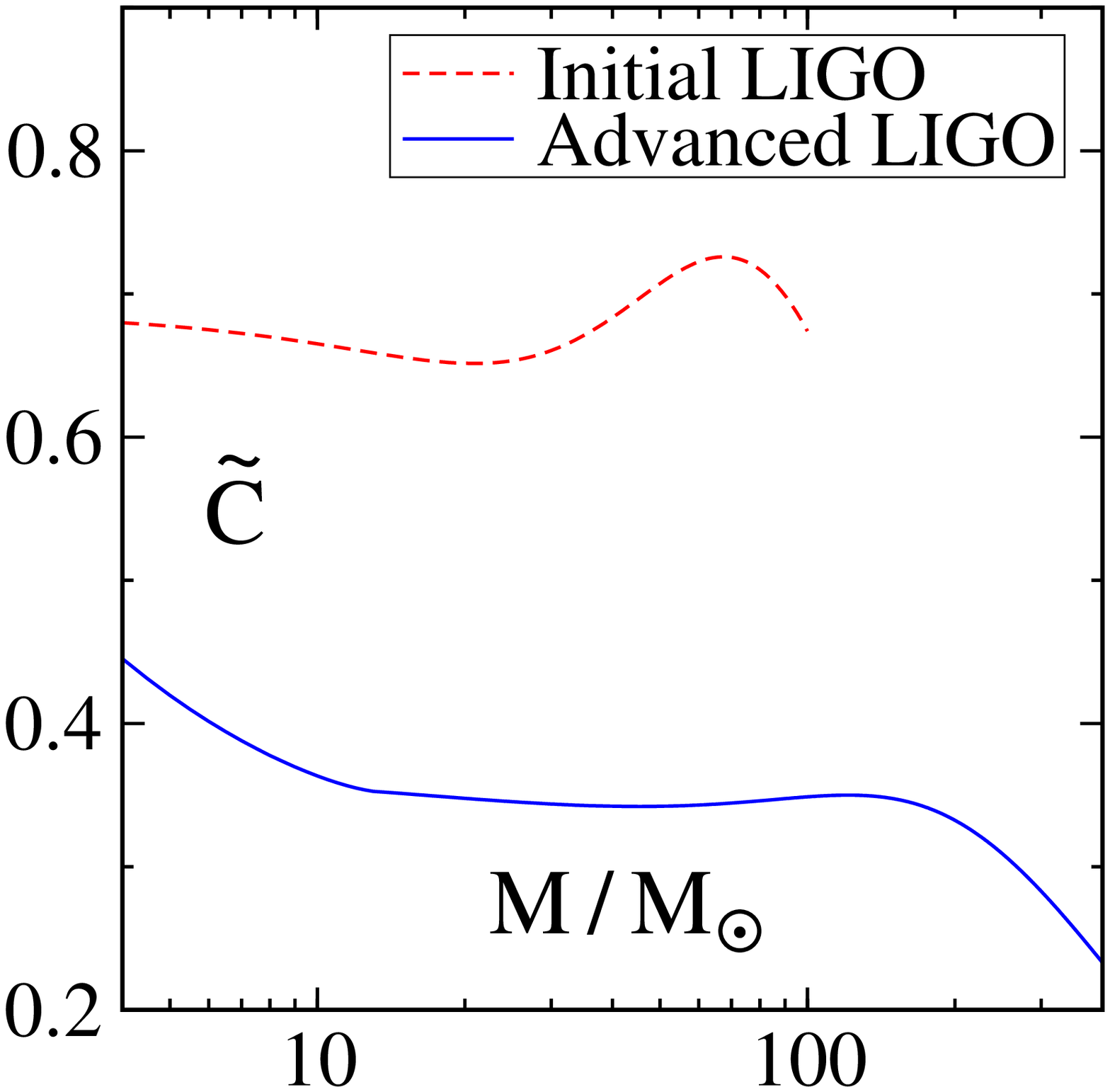}}
\caption{\label{f:CtildeSNratio} Curves illustrate $\widetilde{C}$, the
ratio of the standard signal-to-noise measure $\rho$ to another
non-standard measure defined in Eq.~(\ref{e:ctilde}), as a function of
the total mass for non-spinning equal-mass binary black-hole
waveforms.  Dashed curve is based on the Initial LIGO noise spectrum;
solid curve is based on an Advanced LIGO noise curve.}
\end{figure}

In some circumstances the calibration-accuracy standards obtained
using $\mathrm{max}\left(\delta\chi_R^2+\delta\Phi_R^2\right)$ may be
much stronger than necessary; for example when
$\delta\chi_R^2+\delta\Phi_R^2$ is sharply peaked at frequencies where
the detector noise is large.  In this case it may be advantageous to
employ a different simplification of the accuracy standards.  The
detector calibration-error terms in the noise-weighted norm $\langle
\delta h_R|\delta h_R\rangle$ can also be isolated from the
gravitational-waveform terms with an application of the Schwarz
inequality:
\begin{eqnarray}
&&\!\!\!\!\!\!\!\!\!\!\!\!\!\!\!\!\!\!\!\!\!\!\!
\langle \delta h_R | \delta h_R\rangle =
\int_0^\infty \left(\delta\chi_R^2+\delta\Phi_R^2\right)
\frac{4|h_e|^2}{S_n}df,\nonumber\\
&&\!\!\!\!\!\!\!\!\!\!
\leq \sqrt{\int_0^\infty \frac{4|h_e|^4}{S_n}df}\times\nonumber\\
&&\quad
\left(\sqrt{\int_0^\infty \frac{4\delta\chi_R^4}{S_n}df}
+\sqrt{\int_0^\infty \frac{4\delta\Phi_R^4}{S_n}df}\right).
\label{e:calibrationerror0}
\end{eqnarray}
This inequality can be re-written in the more compact form,
\begin{eqnarray}
\langle\delta h_R | \delta h_R\rangle \leq \frac{\rho^2}{\widetilde{C}^2}
\left(\widetilde{\delta \chi_R}^2+\widetilde{\delta \Phi_R}^2\right),
\label{e:calibrationerror}
\end{eqnarray}
by defining a few useful quantities.  The noise-weighted averages of
$\widetilde{\delta\chi_R}$ and $\widetilde{\delta\Phi_R}$ are defined as
\begin{eqnarray}
\widetilde{\delta\chi_R}^4 &=& \int_0^\infty \delta\chi_R^4
\frac{4\bar n^2}{S_n}df,
\label{e:noiseweightedaverageChi}\\
\widetilde{\delta\Phi_R}^4 &=& \int_0^\infty \delta\Phi_R^4
\frac{4\bar n^2}{S_n}df,
\label{e:noiseweightedaveragePhi}
\end{eqnarray}
where the total detector noise, $\bar n$, is defined as
\begin{eqnarray}
\frac{1}{\bar n^2}=\int_0^\infty \frac{4}{S_n}df.
\end{eqnarray}
Note that the weight, $4 \bar n^2/S_n$, which appears in
Eqs.~(\ref{e:noiseweightedaverageChi}) and
(\ref{e:noiseweightedaveragePhi}) has integral one; so these are true
(noise-weighted) averages of $\delta\chi_R$ and $\delta\Phi_R$.  Note
the unusual fourth power of the averaged quantity which appears in
these definitions.  This is required because that is the power of the
error terms which appear on the right side of
Eq.~(\ref{e:calibrationerror0}).  Finally, the quantity
$\widetilde{C}$ that appears in Eq.~(\ref{e:calibrationerror}) is the
ratio of the standard signal-to-noise measure $\rho$ to another
non-standard measure:
\begin{eqnarray}
\widetilde{C}^{4} 
= \rho^4\left(\int_0^\infty \frac{4|h_e|^4}{\bar n^2 S_n}df\right)^{-1}.
\label{e:ctilde}
\end{eqnarray}
Figure~\ref{f:CtildeSNratio} illustrates $\widetilde{C}$ for
non-spinning equal-mass binary black-hole waveforms using standard
Initial and Advanced LIGO noise
curves~\cite{InitialLIGONoise,AdvancedLIGONoise}. This quantity can be
evaluated in a straightforward way when any class of model waveforms
is computed.  It isn't completely clear why the Advanced LIGO version
of this curve is almost a factor of two smaller than the Initial LIGO
curve.  This may be due to the fact that the integral of $|h_e(f)|^4$
in $\widetilde{C}$, Eq.~(\ref{e:ctilde}), is dominated by its low
frequency contributions where $h_e(f)$ is largest.  The Advanced LIGO
noise curve is significantly smaller in this low-frequency range, so
these contributions are much more significant in that case.  

The maximum calibration-error,
$\mathrm{max}\left(\delta\chi_R^2+\delta\Phi_R^2\right)$, and the
noise-weighted averages, $\widetilde{\delta\chi_R}$ and
$\widetilde{\delta\Phi_R}$, which appear in
Eqs.~(\ref{e:calibrationerror3}) and (\ref{e:calibrationerror})
depend only on information that pertains to the detector itself.  All
of the dependence on the waveform $h_e$ in the original norm,
$\langle\delta h_R|\delta h_R\rangle$, has been moved into the
signal-to-noise ratio $\rho$ and the quantity $\widetilde{C}$.  Thus
the right sides of Eqs.~(\ref{e:calibrationerror3}) and
(\ref{e:calibrationerror}) should be much easier for those performing
detector calibrations to evaluate.  The right sides of
Eqs.~(\ref{e:calibrationerror3}) and (\ref{e:calibrationerror}) are
always larger than the noise-weighted norm $\langle \delta h_R|\delta
h_R\rangle$ that appears on the left.  Sufficient conditions for
model-waveform accuracy based on the maximum calibration error
$\mathrm{max}\left(\delta\chi_R^2+\delta\Phi_R^2\right)$ (or the
noise-weighted averages of the calibration errors
$\widetilde{\delta\chi_R}$ and $\widetilde{\delta\Phi_R}$) can be
obtained therefore by replacing $\langle \delta h_R|\delta h_R\rangle$
with the right side of Eq.~(\ref{e:calibrationerror3}) or
(\ref{e:calibrationerror}) wherever it appears in the accuracy
standards of Eqs.~(\ref{e:measurementlimit}) and
(\ref{e:detectionlimit}).  The resulting accuracy standards based
on the maximum calibration error [and using
the re-written norm of the waveform-modeling error from
Eq.~(\ref{e:time-domain-limit})] become,
\begin{eqnarray}
\frac{1}{C}\frac{||\delta h_m(t)||}{||h_e(t)||}
+\sqrt{\mathrm{max}\left(\delta\chi_R^2+\delta\Phi_R^2\right)} \leq
\frac{1}{\rho},
\label{e:measurementstandard}
\end{eqnarray}
for measurement, and 
\begin{eqnarray}
\frac{1}{C}\frac{||\delta h_m(t)||}{||h_e(t)||}
+\sqrt{\mathrm{max}\left(\delta\chi_R^2+\delta\Phi_R^2\right)}
\leq \sqrt{2\epsilon_\mathrm{max}},
\label{e:detectionstandard}
\end{eqnarray}
for detection.  Analogous versions of the standards based on the
noise-weighted averages of the calibration errors,
$\widetilde{\delta\chi_R}$ and $\widetilde{\delta\Phi_R}$, can be
obtained in a similar way.  Either version of the standards is
sufficient to guarantee the needed level of accuracy for
gravitational-wave data analysis.  The most efficient choice for a
particular detector, and for a particular type of source, will be
determined by whether
$\mathrm{max}\left(\delta\chi_R^2+\delta\Phi_R^2\right)$ or
$(\widetilde{\delta\chi_R}^2+\widetilde{\delta\Phi_R}^2)/\widetilde{C}^2$
is smaller.

\section{Discussion}
\label{s:Discussion}

A new set of accuracy standards have been developed for the calibration
and modeling errors of the waveforms used for gravitational-wave data
analysis.  The basic standards, Eqs.~(\ref{e:measurementlimit}) and
(\ref{e:detectionlimit}), are expressed most naturally in terms of the
noise-weighted inner products commonly used in gravitational-wave data
analysis.  These basic expressions are not very convenient for
actually applying the standards, however. So the basic standards have
been transformed into expressions that are easier to apply:
Eqs.~(\ref{e:measurementstandard}) and (\ref{e:detectionstandard}).
These new representations of the accuracy standards are slightly
stronger, and if satisfied are sufficient to ensure the original
standards are satisfied.

The new accuracy standards, Eqs.~(\ref{e:measurementstandard}) and
(\ref{e:detectionstandard}), prescribe a maximum for the combined
calibration and modeling errors of the gravitational waveforms, not
for each type of error separately.  This means that an increased
accuracy in one allows a somewhat weaker requirement on the other.
Determining how to aportion the accuracy between the two error sources
in an optimal way would require a careful analysis of the costs
involved in reducing the error from each source.  It seems reasonable
to expect that requiring approximately equal accuracy for the
calibration and modeling errors will be close to optimal.  In this
case the transformed expressions for the new accuracy requirements are
\begin{eqnarray}
\frac{1}{C}\frac{||\delta h_m(t)||}{||h_e(t)||}
\approx
\sqrt{\mathrm{max}\left(\delta\chi_R^2+\delta\Phi_R^2\right)}
\lesssim \frac{1}{2\,\rho_\mathrm{max}},
\label{e:measurementstandard2}
\end{eqnarray}
for measurement, and
\begin{eqnarray}
\frac{1}{C}\frac{||\delta h_m(t)||}{||h_e(t)||}
\approx
\sqrt{\mathrm{max}\left(\delta\chi_R^2+\delta\Phi_R^2\right)}
\lesssim \sqrt{\frac{\epsilon_\mathrm{max}}{2}},
\label{e:detectionstandard2}
\end{eqnarray}
for detection; the constant $\rho_\mathrm{max}$ represents the
signal-to-noise ratio of the strongest detected source.  If the
calibration of the instrument must be maintained at the level needed
for accurate measurements of the strongest anticipated sources during
the entire data collection period, then the accuracy requirements on
the waveform-modeling error can be relaxed somewhat for detection:
\begin{eqnarray}
\frac{1}{C}\frac{||\delta h_m(t)||}{||h_e(t)||}
\lesssim \sqrt{2\epsilon_\mathrm{max}}-\frac{1}{2\, 
 \rho_\mathrm{max}}.
\label{e:detectionstandard3}
\end{eqnarray}

These new accuracy standards should be applicable for essentially any
gravitational-wave detector and any type of model waveform used in the
data analysis process.  To apply the standards for each particular
case, the quantities $\rho_\mathrm{max}$, $\epsilon_\mathrm{max}$, $C$
and (perhaps) $\widetilde{C}$ must be evaluated for the particular
family of model waveforms, using the noise spectrum of the particular
detector.  Some insight can be gained into what the standards will
actually look like by examining the case of binary black-hole
inspiral-merger-ringdown waveforms using the Advanced LIGO noise
curve.  The quantities $C$ and $\widetilde{C}$ have been computed for
this case using an equal-mass non-spinning binary black-hole waveform
obtained by matching together numerical and post-Newtonian
waveforms~\cite{Scheel2008,Boyle2008b}.  The results are depicted in
Figs.~\ref{f:CSNratio} and \ref{f:CtildeSNratio} for binary systems
with total masses in the range $4-400M_\odot$.  From these graphs we
see that $C\gtrsim0.019$ and $\widetilde{C}\gtrsim0.23$ for these
waveforms and the Advanced LIGO noise curve.  The strongest binary
black-hole signals in Advanced LIGO are expected to have
signal-to-noise ratios that may be as large as
$\rho_\mathrm{max}\approx 100$.  Assuming the template banks of model
waveforms are constructed in the same way as those for Initial LIGO,
the maximum mismatch compatible with a 10\% event loss rate is
$\epsilon_\mathrm{max}=0.005$.  Substituting these values into the
accuracy standards of Eqs.~(\ref{e:measurementstandard2}) and
(\ref{e:detectionstandard2}) results in the following calibration and
waveform modeling accuracy requirements for Advanced LIGO:
\begin{eqnarray}
53\frac{||\delta h_m(t)||}{||h_e(t)||}
\approx 
\sqrt{\mathrm{max}\left(\delta\chi_R^2+\delta\Phi_R^2\right)}
\lesssim 0.005
\label{e:LIGOmeasurement}
\end{eqnarray}
for measurement, and
\begin{eqnarray}
53\frac{||\delta h_m(t)||}{||h_e(t)||}
\approx 
\sqrt{\mathrm{max}\left(\delta\chi_R^2+\delta\Phi_R^2\right)}
\lesssim 0.05
\label{e:LIGOdetection}
\end{eqnarray}
for detection.  If the calibration accuracy is fixed at the higher
level needed for measurements in the strongest sources for the entire
period in which data is collected, then the standard on
waveform-modeling error for detection can be relaxed to
\begin{eqnarray}
53\frac{||\delta h_m(t)||}{||h_e(t)||}
\lesssim 0.095.
\label{e:LIGOaltdetection}
\end{eqnarray}

A somewhat troubling feature of these conditions is the rather large
coefficient $1/{C}\approx 53$ that multiplies the $L^2$ norms of
$\delta h_m(t)$ in these expressions.  This is really just an artifact
of the extremely long model waveform (containing about 1000 wave
cycles) used here when evaluating $C$~\cite{Lindblom2009b}.  The
quantity $C$ contains the $L^2$ norm of the waveform $h_e(t)$ in its
denominator, and this norm becomes quite large when it is estimated
using model waveforms $h_m$ with many wave cycles.  This issue is
discussed at some length in Ref.~\cite{Lindblom2009b} and will not be
addressed further here, since it does not bear directly on the main
focus of this paper: deriving the optimal levels of calibration error.

\acknowledgments I thank Michael Landry for stimulating my interest in
the questions addressed here and for comments and suggestions for
improving an earlier draft of this paper.  I also think Benjamin Owen
and Yanbei Chen for several helpful discussions on these issues.  This
research was supported in part by a grant from the Sherman Fairchild
Foundation, by NSF grants DMS-0553302, PHY-0601459, and PHY-0652995,
and by NASA grant NNX09AF97G.

\bibliography{../References/References}
\end{document}